PACS 81.05.U-, 63.20.Ry, 63.20.dk, 64.60.A-

# Properties of π-mode vibrations in strained carbon chains

G. M. Chechin[†], D.A. Sizintsev, O.A. Usoltsev

[†]gchechin@gmail.com

Southern Federal University, Department of physics,

Zorge Str., 5, 344090, Rostov-on-Don, Russia

Nonlinear vibrations in strained monoatomic carbon chains are studied with the aid of *ab initio* methods based on the density functional theory. An unexpected phenomenon of structural transformation at the atomic level above a certain value of the strain was revealed in cumulene chain (carbyne-β). This phenomenon is a consequence of stability loss of the old equilibrium atomic positions that occur at small strain, and appearance of two new stable equilibrium positions near each of them. The aforementioned restructuring gives rise to a softening of π-mode whose frequency tends to zero in a certain region of amplitudes when carbon atoms begin to vibrate near new equilibrium positions. This resembles the concept of soft mode whose "freezing" is postulated in the theory of phase transitions in crystals to explain the transitions of displacement type. The dynamical modeling of mass point chains whose particles interact via Lennard-Jones potential can approximate our *ab initio* results well enough. In particular, this study demonstrates an essential role of dipole-dipole interactions between carbon atoms in formation of their new equilibrium positions in the cumulene chain. We believe that computer studying of Lennard-Jones chains enables to predict properties of various dynamical objects in carbon chains (different nonlinear normal modes and their bushes, discrete breathers etc.) which then can be verified by *ab initio* methods.

**Keywords**: carbon chains, nonlinear normal modes, anharmonic vibrations, *ab initio* calculations, phase transitions

## 1. Introduction

Monoatomic carbon chains can exist in two different modifications. The first one is polyyne, or carbyne-α, representing the chain with alternating single and triple bonds [chemical structure $(^{-}C{\equiv}C^{-})_n$]. The second modification is comulene, or carbyne-β, representing the chain with double bonds [chemical structure $(=C=C=)_n$]. Carbyne chains have be claimed to be the strongest material known at the present time. The synthesis of linear carbon chains up to 6000 atoms was reported in 2016. Because of a wide range of interesting properties that were studied or only predicted, such chains may be a potential material for future nanotechnology applications [1, 2].

There are a number of publications devoted to study mechanic and electronic properties of carbyne chains with the aid of *ab initio* methods based on the density functional theory (DFT). For example, let us refer to the paper [3]. DFT *ab initio* methods allow the authors of this paper to obtain a number of interesting results for strained carbon chains. They revealed that distribution of bond length and magnetic moments at atomic sites exhibit even-odd disparity depending on the number of carbon atoms in the chain and on the type of saturation of these atoms at both ends. It was also found that a local perturbation created by a small displacement of the single carbon atom at the center of a long chain induces oscillations of atomic forces and charge density, which are carried to long distances over the chain.

In the present paper, we study *nonlinear vibrations* of carbon atoms in cumulene described by π-mode, which is one of the nonlinear normal modes by Rosenberg (NNMs) [4].

## 2. Symmetry-determined NNMs

Nonlinear normal mode by Rosenberg represents a periodic vibrational regime for which all degrees of freedom $x_i(t)$, at any time $t$, are proportional to each other (in the coordinate space, it corresponds to a straight line). Mathematically, this definition can be written in the following form:

$$x_i(t) = a_i f(t)(i = 1..N). \tag{1}$$

Here $a_i$ are constant coefficients, while $f(t)$ is the same time-dependent function for all degrees of freedom. If we know the explicit form of dynamical equations, the substitution of ansatz (1) into these equations leads to a system of ($N$-$1$) nonlinear algebraic (possibly transcendental) equations for coefficients $a_i$ and a single differential equation for the function $f(t)$. Note that the conventional linear normal modes represent a special case of Rosenberg's modes. In this case,

$$f(t) = cos(\omega t + \varphi_0),$$

where $\omega$ and $\varphi_0$ - are frequency and initial phase, while coefficients $\{a_i \mid i = 1..N\}$ are particle oscillation amplitudes. Each NNM describes a *periodic* oscillation and, in contrast to the case of linear normal modes, the number of such modes has no relation with the dimension of the system. Unfortunately, NNMs by Rosenberg can exist only in very specific dynamical systems, in particular, in those, whose potential energy is a homogeneous function of all its arguments.

On the other hand, it was shown in [5, 6] that there can be some *symmetry-related* reasons for existence of NNMs in systems with arbitrary interparticle interactions. These dynamical objects we call Rosenberg's symmetry-related nonlinear normal modes (bellow the specification "symmetry-related" is omitted, because we consider only such type of modes). In the above papers, it was also proved that in the monoatomic chains (with periodic boundary conditions) which are described by the symmetry group $D_N$ there can exist only three NNMs: π-, σ- and τ-mode. The unit cell of the chain vibrational state is larger than that of the equilibrium state by 2, 3 or 4 times for π-, σ- and τ-modes, respectively. Thus, these modes can be excited only in the chains that consist of $N$ particles, divisible by 2 for π-mode, by 3 for σ-mode and by 4 for τ-mode. The π-mode is the most short-wave mode. Any pair of adjacent particles in the corresponding dynamical regime oscillate in antiphase with the same amplitude. Thus, the vibrational state of the chain at any time *t* can be described by the following set of atomic displacements:

$$[a(t), - a(t) \mid a(t), - a(t) \mid \ldots \mid a(t), - a(t)].$$

We rewrite this set in the brief form [*a*, -*a*], indicating atomic displacements only in one unit cell of the vibrational state (this cell contains two carbon atoms) and omitting the time argument *t*.

Similarly, for σ-mode we have the displacement pattern of the form [*a*, 0, -*a*] (*N mod* 3 = 0), while for τ-mode it can be written as [*a*, 0, -*a*, 0] (*N mod* 4 = 0). All above discussed NNMs represent one-parametric dynamical regimes, since their atomic patterns depend on the single parameter *a*.

The existence of only a *finite* number of NNMs in the chain was proved in [5, 6] with the aid of specific group-theoretical methods. Vibrational states with unit cell whose size larger than that of π-, σ- and τ-modes represent *quasi-periodic* dynamical regimes which are *m*-dimensional (*m*>1) *bushes* of NNMs [7, 8] (they describe quasi-periodic motion with *m* fundamental frequencies in the Fourier spectrum).

## 3. Amplitude-frequency characteristics of non-linear vibrations of carbon atoms in π-mode

We investigate *longitudinal* atomic vibrations of *uniformly strained* carbon chains in π-mode dynamical regime. This strain is modeled by an artificial increase of the unit cell size ($R$) with respect to that of the chain without any strain ($R_0$). Thus, speaking about the strain of the chain by $\eta$ per cent, we mean that $R = R_0\,(1+ \eta)$.

As was already discussed, the unit cell for describing π-mode vibrations of the chain with periodic boundary conditions is twice larger than that of equilibrium state. Since two carbon atoms in this unit cell possess, at any time *t*, displacements $x(t)$ and $-x(t)$, one can discuss the time evolution of only one of them choosing the origin at its equilibrium position. To excite π-mode vibrations in the chain we assume $x(0) = a$, $\dot{x}(0) = 0$.

An important feature of nonlinear vibrations is the dependence $\omega(a)$ of the frequency ($\omega$) on the amplitude ($a$). To find this dependence, we carried out a series *ab initio* calculations based on the density functional theory [9, 10, 11] using the software package ABINIT [12, 13]. The Born-Oppenheimer approximation was used to separate fast motion of electrons and slow motion of nuclei. At each time step for fixed positions of nuclei, self-consistent electron density distribution is calculated (by solving Kohn-Sham quantum-mechanical equations). Then forces acting on the nuclei are computed, and their new configuration is found by using one step of solution of classical dynamical equations. For this new configuration, the procedure of self-consistency for the electronic subsystem is repeated.

All calculations were carried out in the framework of the local density approximation (LDA). Pseudo-potential by Troullier-Martins was used to describe the field of the carbon atoms inner shells in the process of the Kohn-Sham equations solving with the aid of plane waves basis (with energy cutoff equal to 1360 eV). The convergence for energy is chosen as $10^{-8}$ eV between two steps.

For each computational run, with fixed values of the chain strain and the amplitude (parameter) $a$ of π-mode, the frequency $\omega(a)$ was calculated. In Fig.1, we present the function $\omega(a)$ for chains with 5%, 7.5% and 10% strain. This figure demonstrates that *hard type of nonlinearity* appears at relatively small strains (increase of amplitude entails increasing frequency).

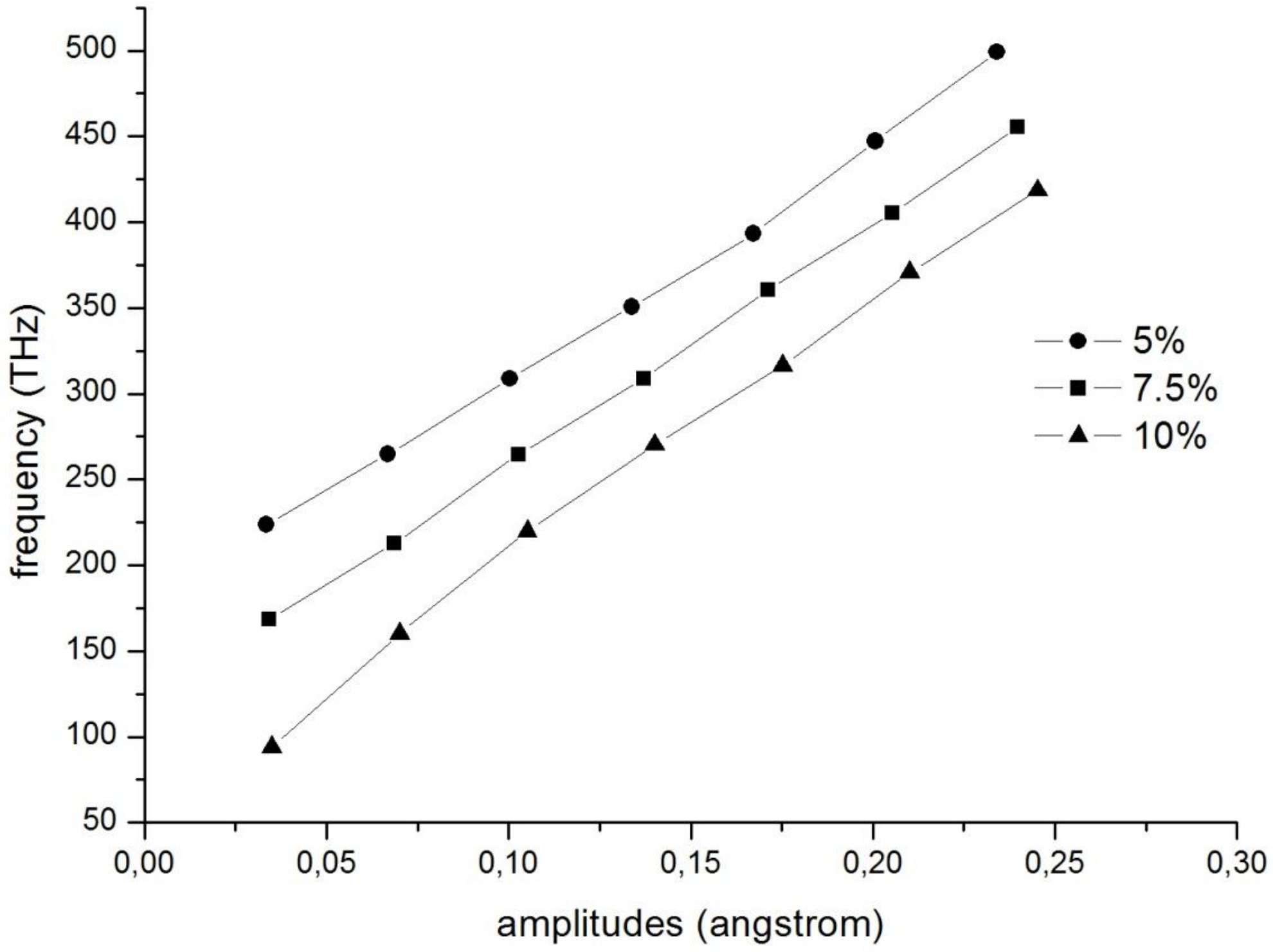

**Fig.1.** Dependence of the frequency (ω) of π-mode on its amplitude (*a*) for small strains of carbyne chain.

However, we revealed unexpected behavior of the function $\omega(a)$ for 15% strain! Indeed, one can see in Fig.2 that this function turns out to be nonmonotonic, and *softening* of π-mode, at a certain interval of the parameter (*a*), takes place (up to point **A** and after point **B** hard type of nonlinearity occurs, while soft nonlinearity is observed between points **A** and **B**).

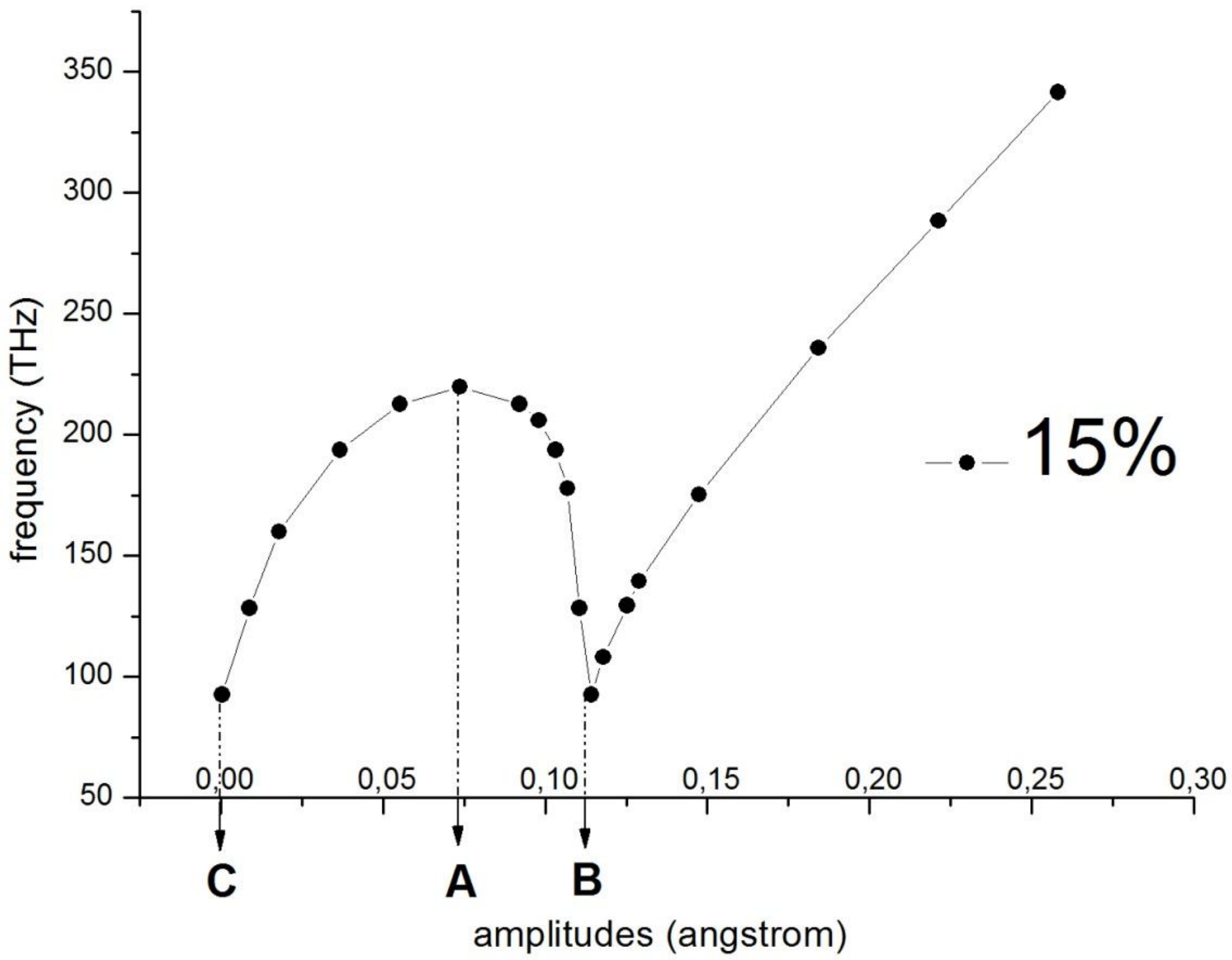

**Fig.2.** Dependence of the frequency (ω) of π-mode on its amplitude (*a*) for 15% strain of carbyne chain.

Phenomenon of π-mode softening in the strained carbyne seems to be very interesting. Indeed, in the framework of the phase transitions theory, there is well-known concept of *soft modes* whose "freezing" induce transitions of displacement type. In this case, one usually refer to some phenomenological arguments about properties of electron-phonon interactions depending on external conditions such as temperature, pressure etc. In contrast, the softening of π-mode that we revealed appears in *ab initio* calculations *without* any additional assumptions of phenomenological nature.

It is essential that for the π-mode parameter *a* belonging to the interval [**C**, **B**] (Fig.2) carbon atom oscillates about a *new equilibrium position* different from the old one at *x*=0. This effect is illustrated in Fig.3 for the chain strained by 15% (detail discussion of these oscillations is given bellow in Sec.4).

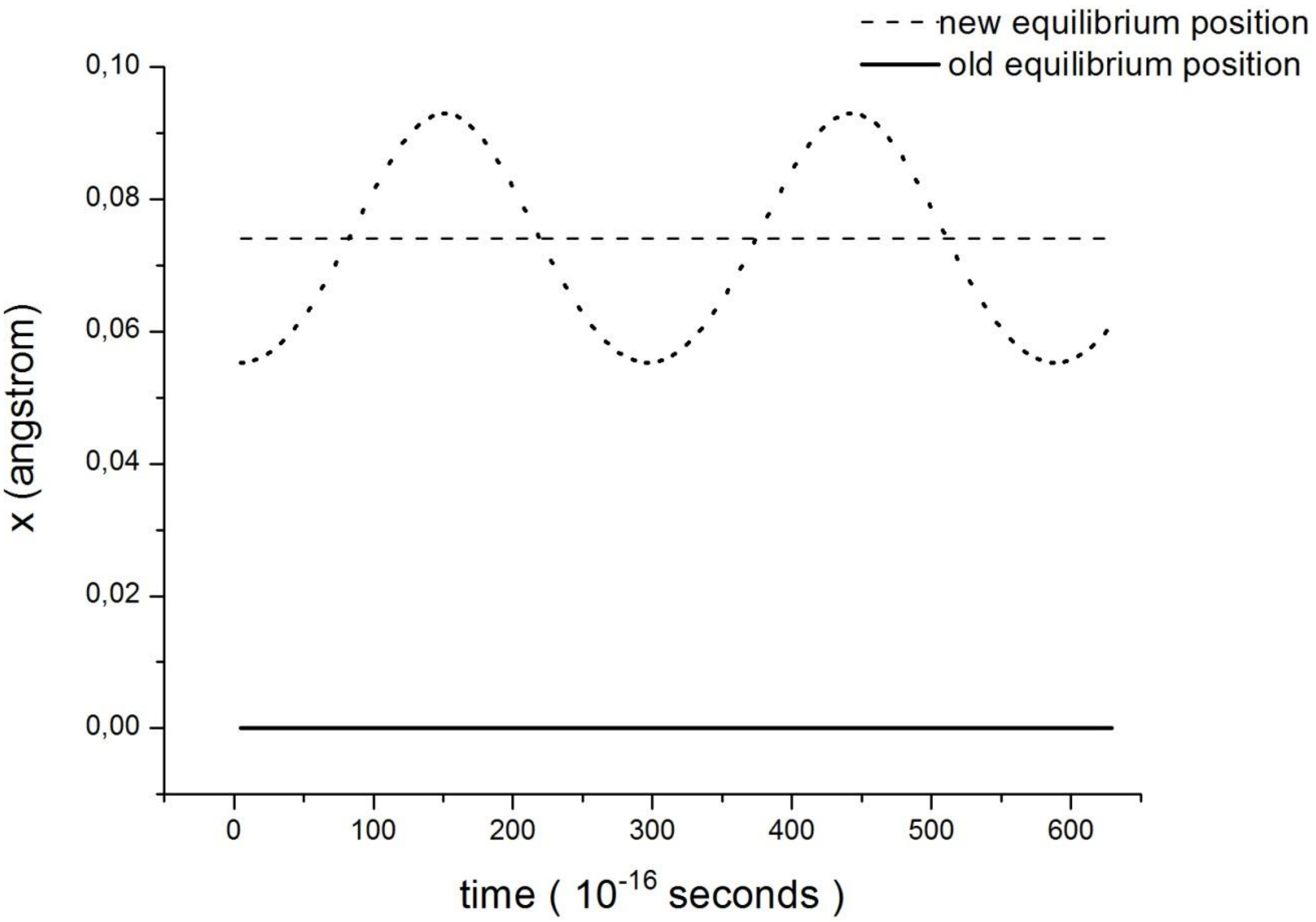

**(a)**

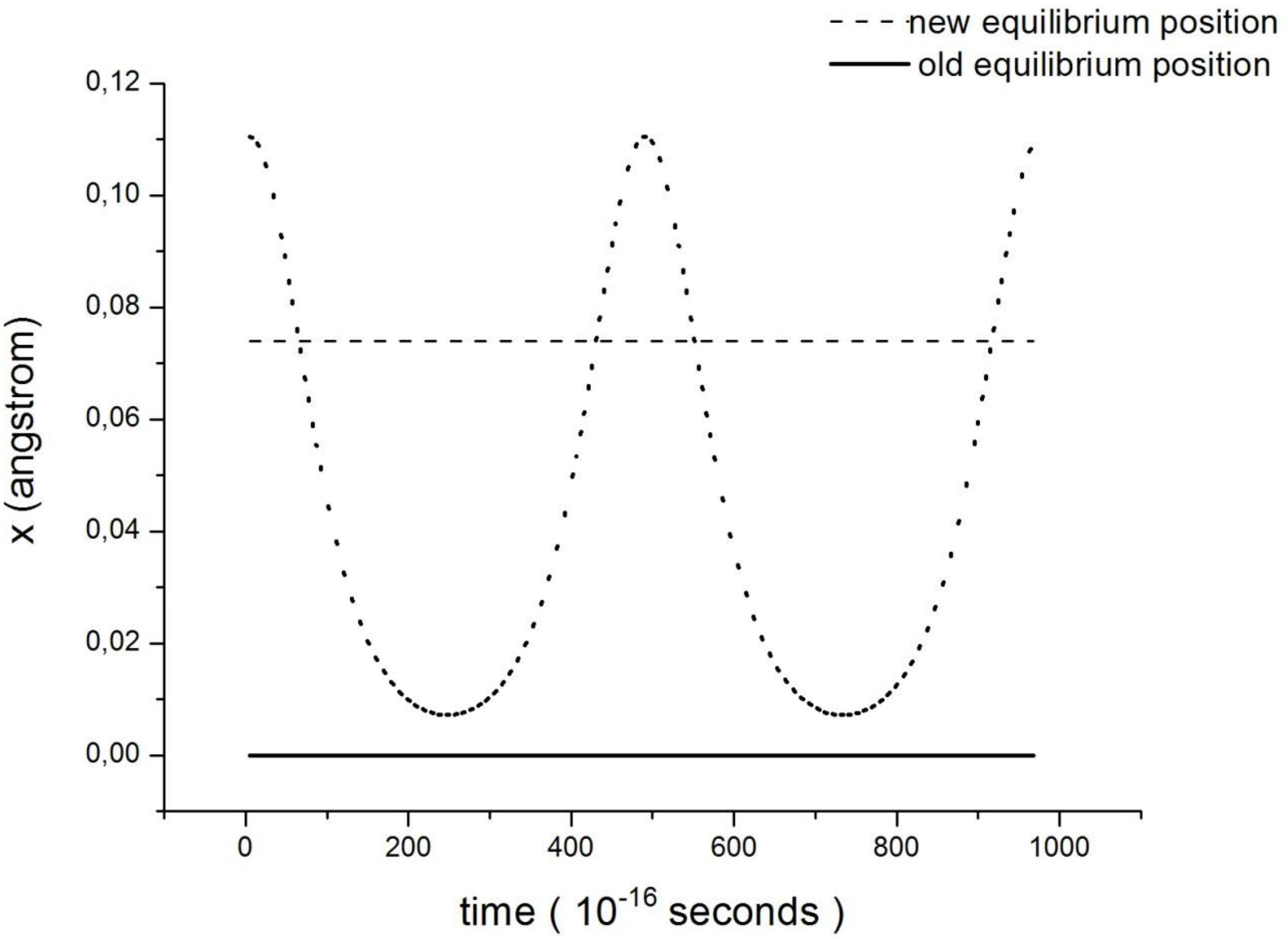

**(b)**

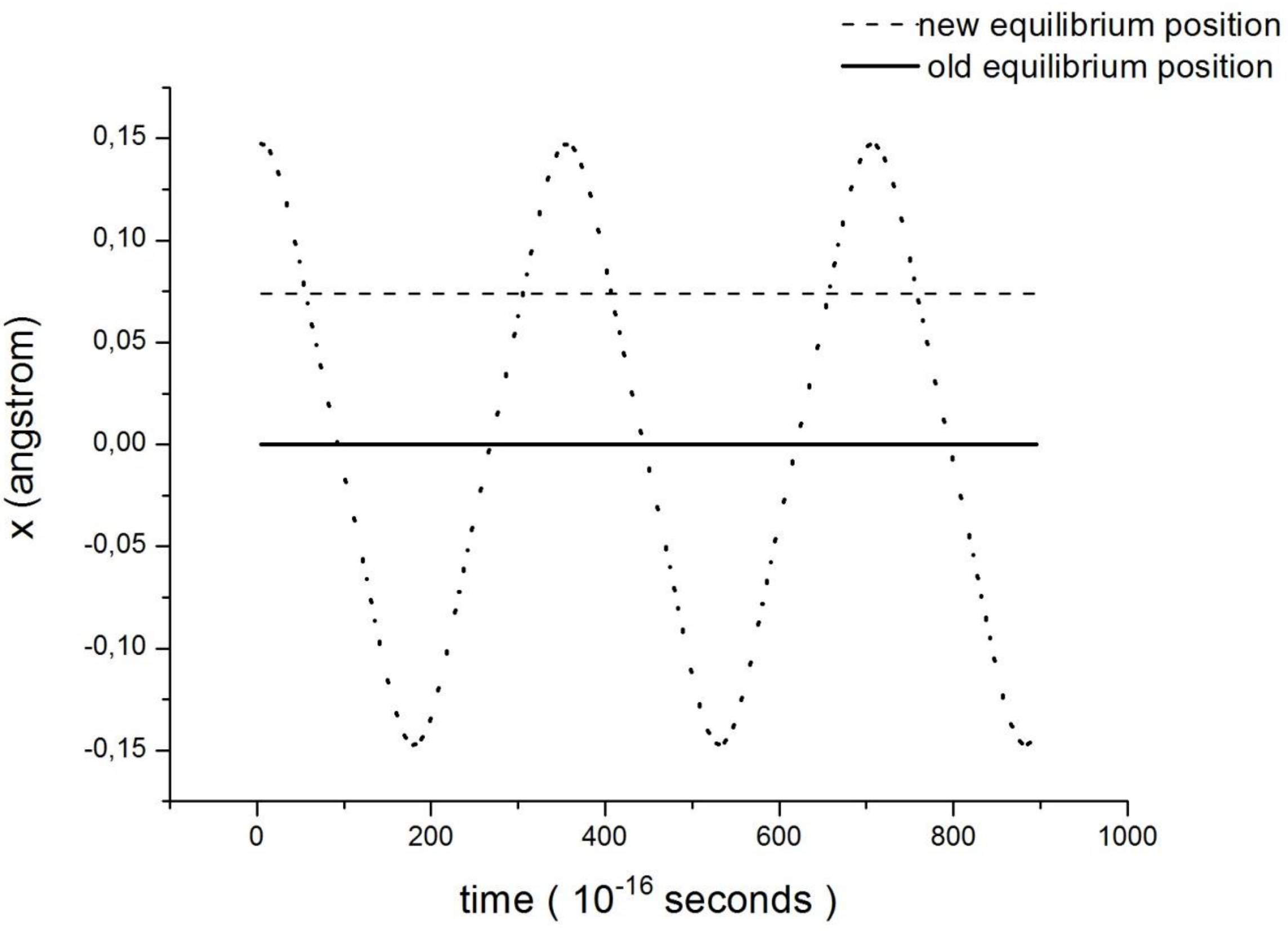

**(c)**

**Fig.3.** Oscillations of the carbon atom for different π-mode amplitudes:
**(a)** oscillations in the small potential well near the new equilibrium position;
**(b)** oscillations before the escape from the small potential well; **(c)** oscillations in the large potential well with respect to the old equilibrium position.

To explain the above behavior of the function $\omega(a)$ one can study the potential energy $U(a)$ as a function of the π-mode amplitude $a$ for different strains of the carbon chain.

## 4. Energy of π-mode in strained carbon chains

We fix the configuration of carbon nuclei choosing a concrete value of π-mode amplitude $a$. Using the software package ABINIT, we then find potential energy $U(a)$ of this configuration[1].

The energy profiles $U(a)$ for different strains of carbon chains are shown in Fig.4. From this figure, one can see that with increase of the strain, the plots $U(a)$ become flatter near the origin ($a = 0$) and their curvature changes sign after passing through zero. As a result, the *old* equilibrium position ($a = 0$) becomes unstable, and two *new* minima appear at equal distances on both sides of the origin. These new equilibrium positions turn out to be stable. This specific behavior of potential energy of strained carbon chains help us to explain the properties of amplitude-frequency dependence $\omega(a)$ shown in Fig.2. Let us consider this question in more detail.

---

[1] We excite π-mode vibrations assigning all carbon atoms the same displacements ($a$) and zero velocities at initial time. Thus, the total energy $E(a)$ of the carbyne chain at this instant is equal to its potential energy $U(a)$.

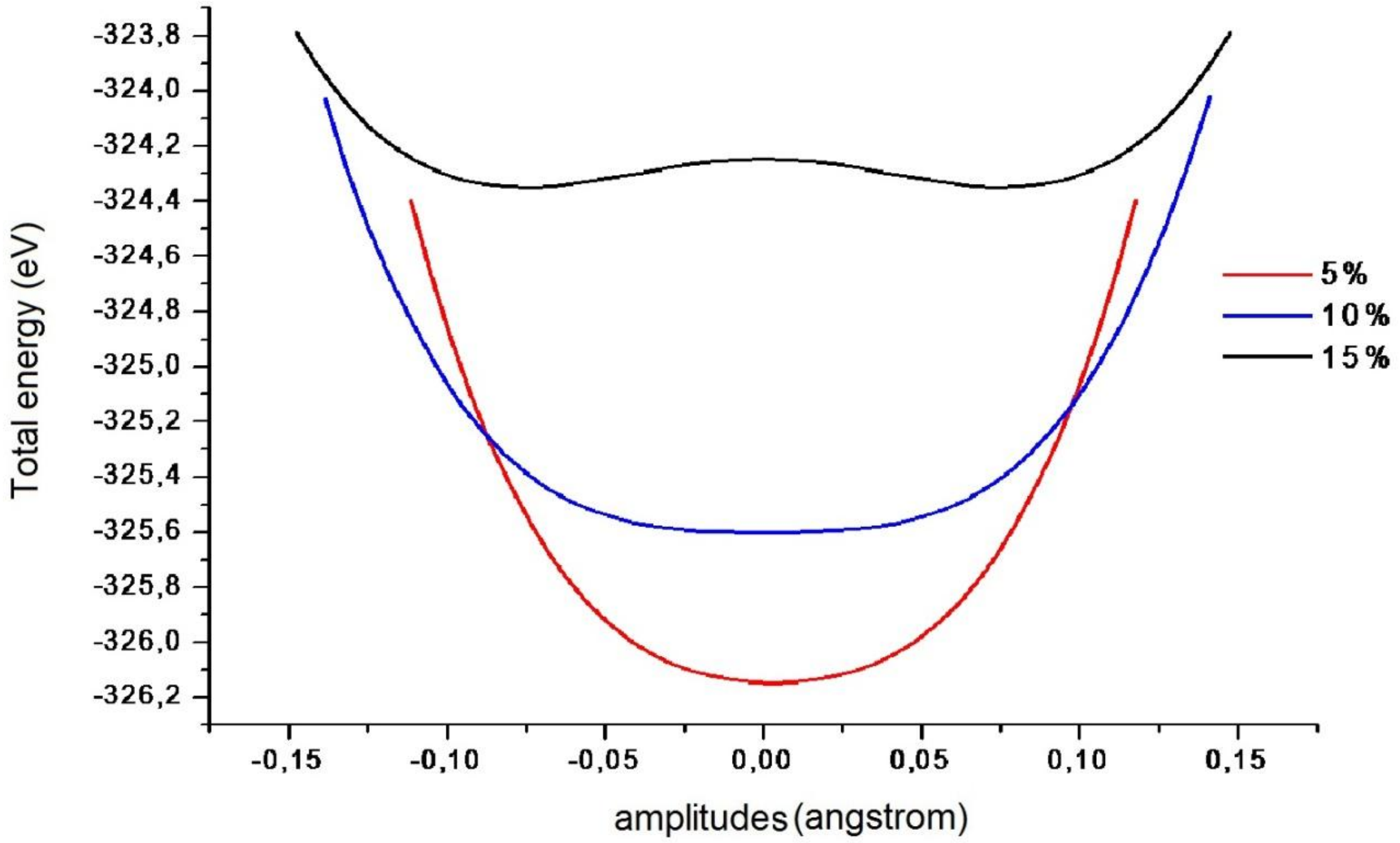

**Fig.4.** Dependence of the potential plots of carbyne chains for different strains.

Point C in Fig.5 corresponds to the old equilibrium position ($a$ = 0) that now becomes unstable. For arbitrarily small positive values of π-mode amplitude $a$, carbon atom begins to vibrate almost from *zero frequency* in the right potential well about new equilibrium position corresponding to its bottom (point A in Fig.5). Indeed, in the absence of any small perturbations, carbon atom, being placed on the hilltop (point **C**), will be located at this point infinitely long time and, therefore, its "frequency" turns out to be zero. Since such situation is not possible, we must consider only small (e.g., positive) values of the parameter $a$. The corresponding period of oscillations should be sufficiently large, and therefore, the frequency will be small enough.

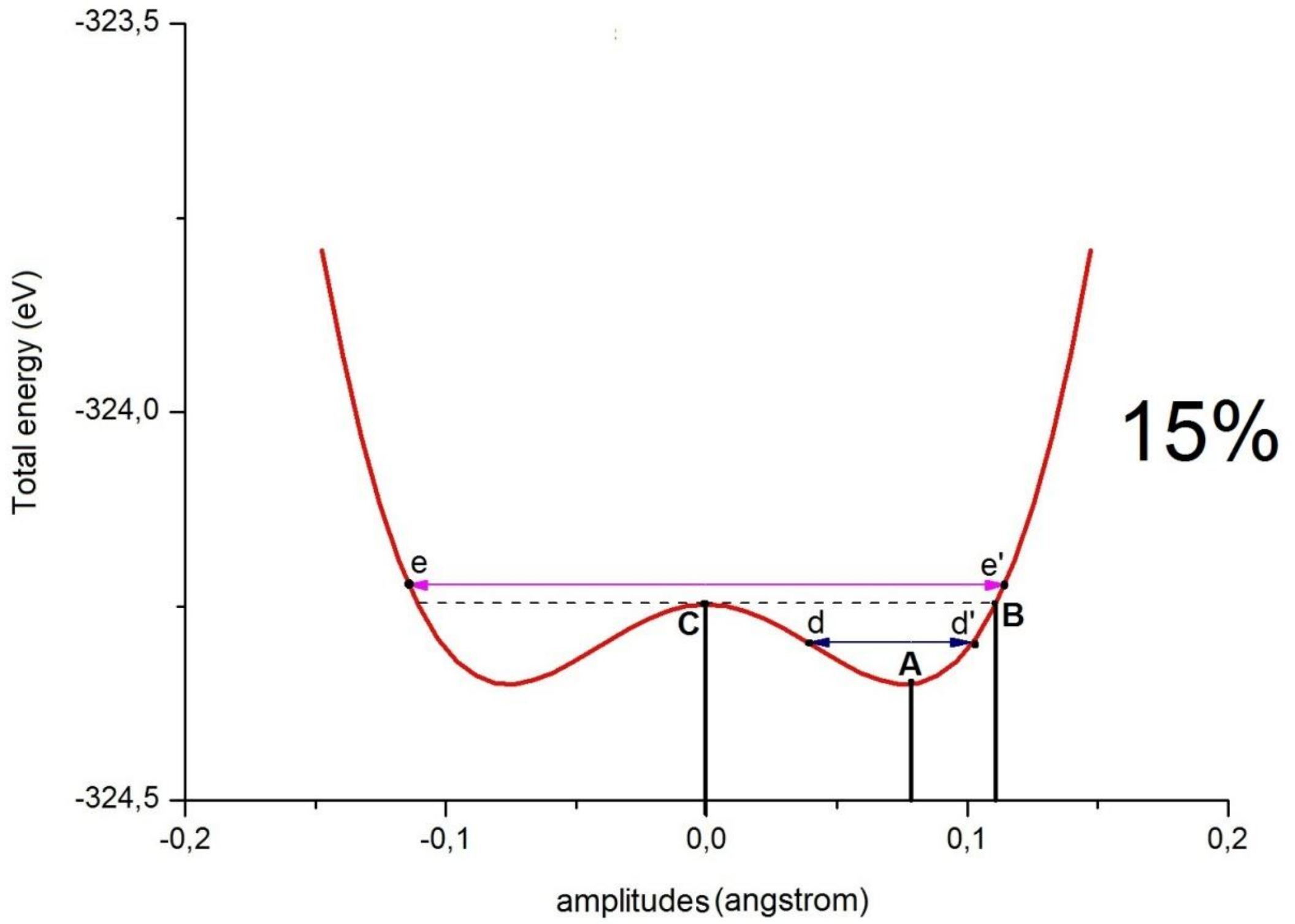

**Fig.5.** Carbyne potential plot for 15% strain and π-mode oscillations of carbon atom with respect to new (point A) and old (point C) equilibrium positions.

As can be seen from Fig. 5, a further increase of the parameter *a* leads to the energy *decrease*, and our atom begins to move between two turning points of the right potential well that are indicated by symbols d and d’. In turn, this means *decrease* of the atomic vibrational amplitude and *increase* of the frequency which reaches its maximal values near the bottom of the well (point A).

Continue to increase the π-mode parameter (*a*), one can note, that the vibrational amplitude begins to increase, while the frequency decreases and runs through the same values that take place at the previous stage of its increasing (one obtains the *same* frequency starting from both points d and d’).

As we approach the point B, the frequency ($\omega$) again tends to zero. Further increase of the π -mode amplitude *a* leads to the qualitative change of vibrations: the carbon atom begin to oscillate in *large well* with respect to the old equilibrium position $x$=0 (for example, these oscillations occur between points e and e’). Thus, at the exit from the right potential well, the function $\omega(a)$ changes abruptly from zero to a certain finite value.

Above discussed oscillations of the carbon atom are presented in Fig.3. The plots 3a and 3b show oscillations of this atom for $a$=0.055 Å and $a$= 0.111 Å with respect to the new equilibrium position $x$=0.074 Å in the small right potential well, while plot 3c corresponds to the oscillations by the old equilibrium position at $x$=0 ($a$=0.147 Å) in the large potential well.

Let us note that both potential wells, left and right, are not symmetric with respect to their bottoms. This is the cause why the plot $\omega(a)$ in Fig.2 turns out to be asymmetric relative to the hilltop of this function.

## 5. Studying of carbyne chain vibrations in the framework of molecular dynamics

The main idea of the molecular dynamics can be formulated as follows. Molecules (atoms) are replaced by mass points whose interactions are described with the aid of some phenomenological potentials. For the obtained dynamical system, equations of classical mechanics are solved. In the framework of quantum mechanics, such approach cannot be considered as sufficiently adequate, because it is difficult or impossible to find potentials which are good enough to take into account the influence of atomic electron shells on dynamical properties of the original physical system. That is why one has to use very complicated many-particle potentials which possess different forms for different geometry of the interacting atoms and contain phenomenological constants defined by huge tables (see, for example, [15]).

However, in some cases, one can obtain reasonable results even with the aid of simple pair potentials, such as those by Morse, Lennard-Jones etc. For example, one can refer to the paper [14] devoted to study discrete breathers in 2D and 3D Morse crystals. Bellow, we try to explain above-discussed results, obtained by *ab* *initio* calculations for π-mode dynamics in carbyne, using the model of a chain whose particles interact via Lennard-Jones potential (LJ-chain).

Let us remind some well-known properties of this potential which can be written in the form

$$\varphi(r) = \frac{A}{r^{12}} - \frac{B}{r^{6}}\,. \qquad (2)$$

Here the first term describes repulsion between two particles that are at distance r from each other, while the second term describes their attraction. The space dependence of this attraction can be explained in the framework of quantum mechanics as a result of the induced dipole-dipole interaction, while the 12-th degree of the distance r in repulsive term of Eq.(2) is introduced only for computational convenience. An important feature of the Lennard-Jones potential is that its both constants, A and B, can be chosen equal to unity (A=B=1) without loss of generality, if we make an appropriate scaling of space and time variables in dynamical equations constructing with the aid of this potential.

In Fig.6, we present the dependence of π-mode energy on its amplitude *a*. The solid line corresponds to the results of *ab initio* calculations, while dot and dash-dot lines were obtained for the chains whose particles interact via Lennard-Jones and Morse potential, respectively. It can be seen from this figure, that the energy calculated by *ab initio* methods for amplitudes *a* in the interval [-0.125, 0.125] is approximated well enough by that obtained for the LJ-chain (mean square deviation of these two energy graphs is equal to 0.026 eV).

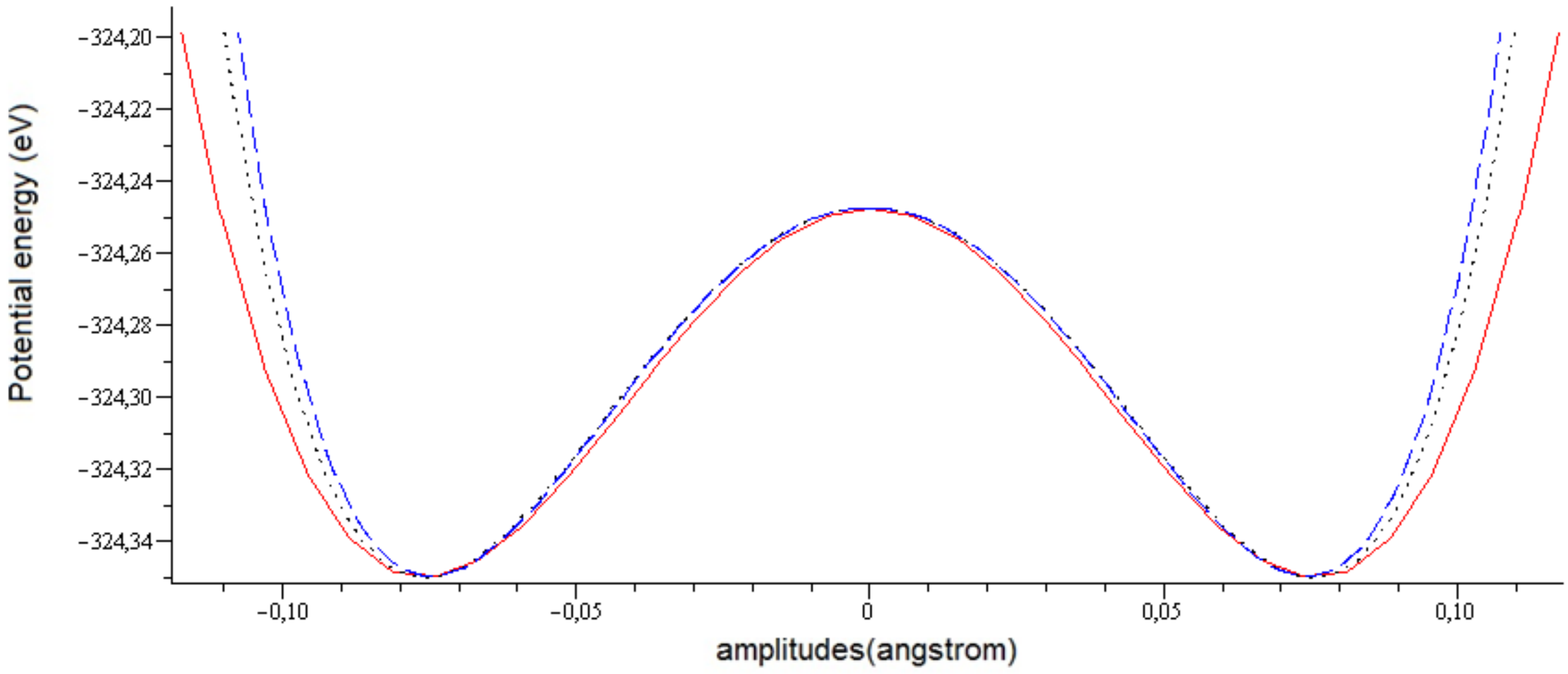

**Fig.6.** Potential plots for carbyne chain: *ab initio* results (solid line), Morse chain (dotted line) and Lennard-Jones chain (dash line).

We verified that dynamical properties of the π-mode obtained by *ab initio* study and by molecular dynamics for LJ-chain are also in a sufficiently good agreement.

Note that such a relationship between results of the carbyne *ab initio* study and that of molecular dynamics modeling for LJ-chains takes place not only for 15% strain, but for all other strains of the carbyne chains considered in this paper.

The above facts seem to be very important. Indeed, computer experiments based on the density functional theory require very long time in contrast to those based on the methods of molecular dynamics. Therefore, one can use the LJ-chain modeling to predict some dynamical properties of the carbyne which then may be verified by *ab initio* calculations.

## 6. Simple interpretation of vibrational properties of strained carbyne chains

It is very interesting to clear up the nature of evolution of the carbyne properties with increasing of its strain. Why old equilibrium positions lose stability and new stable equilibrium positions appear? This problem can be understood with the aid of the following simple model.

Let us consider a system of three carbon atoms located at a distance R from each other symmetrically about the origin x=0 (see Fig.7). Positions of atoms 1 and 3 are fixed, while the "inner" atom 2 we displace by a certain value $x$ from its "old" equilibrium position at origin. We assume that atoms interact via Lennard-Jones potential $\varphi(x)$ from Eq.(2) with A=B=1.

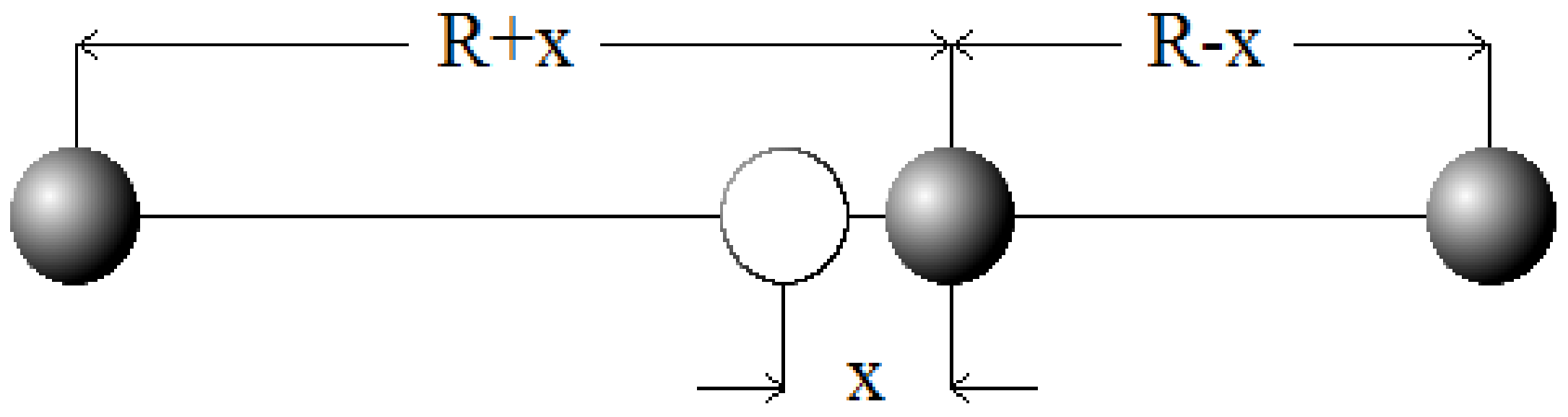

**Fig.7.** Simple model of 3 atoms interacting via Lennard-Jones potential. Atoms 1, 3 are fixed, while atom 2 is displaced by $x$ from its old equilibrium position in origin.

If atom 2 gets a new equilibrium position at $x\neq0$, then the forces $f(r)$ acting on it from the left and right neighbors must be equal:

$$f(R+x) = f(R-x). \qquad (3)$$

Hear

$$f(r) = -\frac{d\varphi}{dr} = \frac{1}{r^{13}} - \frac{1}{r^{7}}. \qquad (4)$$

Eq.(3) represents a nonlinear equation with respect to x that may have *several* real roots. We depict this situation in Fig. (8) where *two* intersection points of the functions $f(R+x)$ and $f(R-x)$ at $x$=0 and at $x$=0.130 take place. Here R=1.291 (this value corresponds to 15% stretching of the carbyne chain).

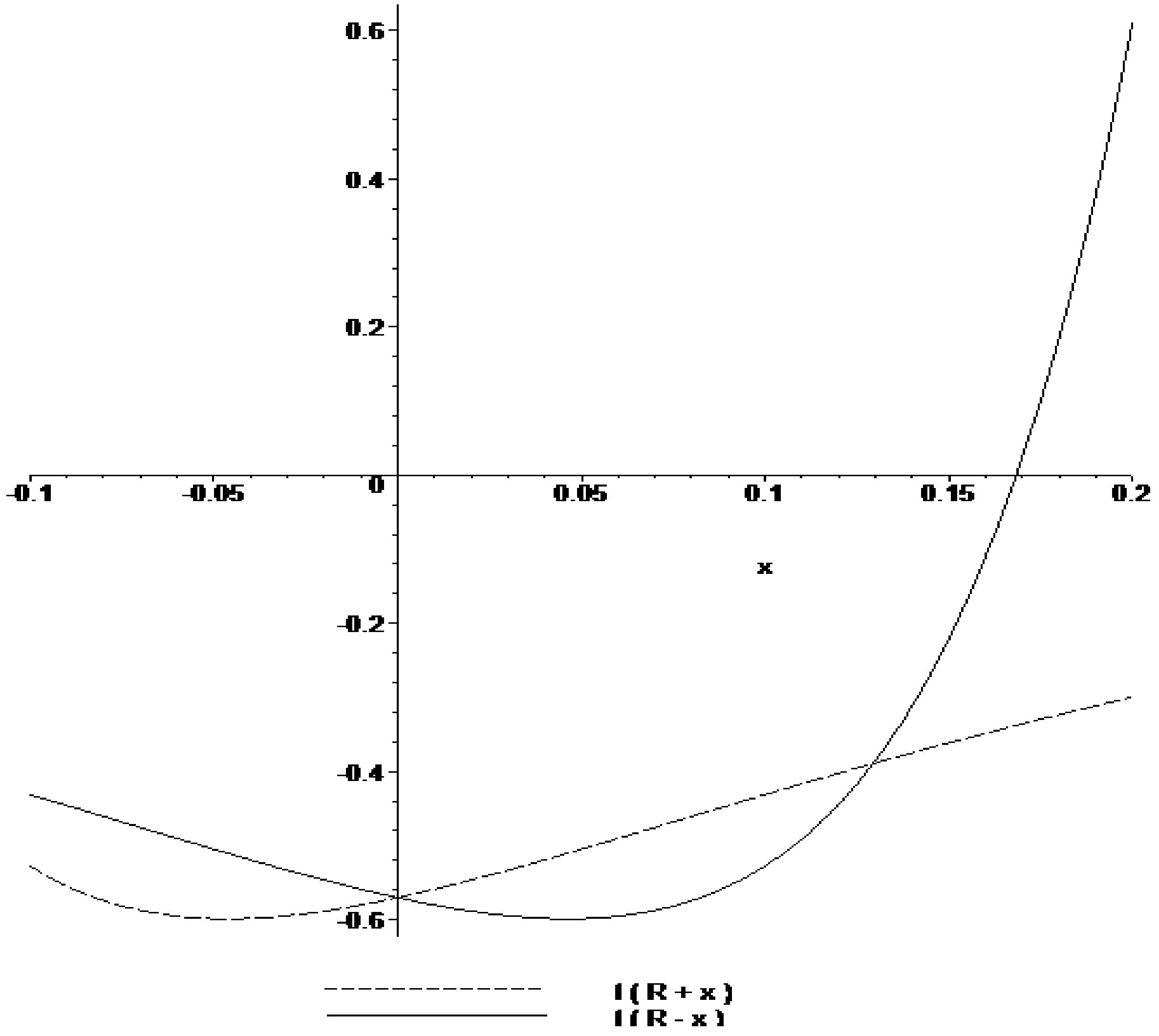

**Fig.8.** Appearance of a new equilibrium position in the simple model of three particles interacting via Lennard-Jones potential.

The root $x$=0 is associated with old equilibrium position that becomes unstable, while the root $x$=0.130 is associated with the new equilibrium position which turns out to be stable.

It is also easy to verify that the curvature of the potential energy of the considered sistem is negative for R=1.20 (stable equilibrium), while it becomes positive for R=1.25 (unstable equilibrium).

Certainly, one can reveal the similar behavior of the potential energy considering LJ-chains with a large number of atoms.

## 7. Conclusion

Let us summarize the above presented results.

We study nonlinear atomic vibrations in strained cumulene chains described by π-mode (it is one of three symmetry-determined nonlinear normal modes by Rosenberg that can be excited in such chains under periodic boundary conditions).With the aid of *ab initio* methods based on the density functional theory (DFT), we have revealed the π-mode *softening* in a certain interval of its amplitudes. This phenomenon is well-known in the theory of structural phase transitions in crystals. Indeed, transitions of displacement type are often treated in the framework of the concept of the vibrational soft mode *freezing*. As a rule, this idea is tried to justify by some phenomenological arguments. Let us emphasize that in our *ab initio* simulation π-mode softening arose *without* any additional assumptions.

Analysis of the carbyne potential energy in the vicinity of the π-mode softening region shows that old equilibrium positions of the carbon atoms *lose stability* and two *new* stable equilibrium positions appear near each of them. Thus, the essential transformation of the carbyne structure at the atomic level takes place under the action of a certain uniform strain. This phenomenon of π-mode softening can be then explained by the fact that carbon atoms begin to oscillate in potential wells near the new equilibrium positions.

It was also found that chains of mass points interacting via Lennard-Jones potential can demonstrate, under the appropriate strain, properties similar to those of carbyne which have been revealed with the aid of DFT *ab initio* calculations. We hope that computer modeling of such Lennard-Jones chains enables to predict properties of various dynamical objects in carbon chains (different nonlinear normal modes and their bushes, discrete breathers etc.) which then can be verified by *ab initio* methods. Results of these studies will be published elsewhere.

*The authors are sincerely grateful to Profs. V. P. Sakhnenko and S. V. Dmitriev for useful discussions, and to I. P. Lobzenko for assistance in application of the software package ABINIT. This work was supported by the Russian Science Federation (Grant No. 14-13-00982).*